

\def\singlespace{\normalbaselines}
\def\oneandahalfspace{\baselineskip=1.15\normalbaselineskip plus 1pt
\lineskip=2pt\lineskiplimit=1pt}

\def\np{\vfill\eject}
\def\nl{\hfil\break}

\def\nofirstpagenoten{\nopagenumbers\footline={\ifnum\pageno>1\tenrm
\hss\folio\hss\fi}}
\def\nofirstpagenotwelve{\nopagenumbers\footline={\ifnum\pageno>1\twelverm
\hss\folio\hss\fi}}
\def\leaderfill{\leaders\hbox to 1em{\hss.\hss}\hfill}
\def\ft#1#2{{\textstyle{{#1}\over{#2}}}}
\def\frac#1/#2{\leavevmode\kern.1em
\raise.5ex\hbox{\the\scriptfont0 #1}\kern-.1em/\kern-.15em
\lower.25ex\hbox{\the\scriptfont0 #2}}
\def\sfrac#1/#2{\leavevmode\kern.1em
\raise.5ex\hbox{\the\scriptscriptfont0 #1}\kern-.1em/\kern-.15em
\lower.25ex\hbox{\the\scriptscriptfont0 #2}}


\parindent=20pt
\def\narrow{\advance\leftskip by 40pt \advance\rightskip by 40pt}

\def\AB{\bigskip
        \centerline{\bf ABSTRACT}\medskip\narrow}
\def\nonarrower{\advance\leftskip by -40pt\advance\rightskip by -40pt}
\def\AE{\bigskip\nonarrower}

\def\boxit#1{\vbox{\hrule\hbox{\vrule\kern3pt
        \vbox{\kern3pt#1\kern3pt}\kern3pt\vrule}\hrule}}

\def\gtorder{\mathrel{\raise.3ex\hbox{$>$}\mkern-14mu
             \lower0.6ex\hbox{$\sim$}}}
\def\ltorder{\mathrel{\raise.3ex\hbox{$<$}|mkern-14mu
             \lower0.6ex\hbox{\sim$}}}
\def\dalemb#1#2{{\vbox{\hrule height .#2pt
        \hbox{\vrule width.#2pt height#1pt \kern#1pt
                \vrule width.#2pt}
        \hrule height.#2pt}}}

\font\fourteentt=cmtt10 scaled \magstep2
\font\fourteenbf=cmbx12 scaled \magstep1
\font\fourteenrm=cmr12 scaled \magstep1
\font\fourteeni=cmmi12 scaled \magstep1
\font\fourteenss=cmss12 scaled \magstep1
\font\fourteensy=cmsy10 scaled \magstep2
\font\fourteensl=cmsl12 scaled \magstep1
\font\fourteenex=cmex10 scaled \magstep2
\font\fourteenit=cmti12 scaled \magstep1
\font\twelvett=cmtt10 scaled \magstep1 \font\twelvebf=cmbx12
\font\twelverm=cmr12 \font\twelvei=cmmi12
\font\twelvess=cmss12 \font\twelvesy=cmsy10 scaled \magstep1
\font\twelvesl=cmsl12 \font\twelveex=cmex10 scaled \magstep1
\font\twelveit=cmti12
\font\tenss=cmss10
 
 \font\ninebf=cmbx7 scaled \magstep1
\font\ninerm=cmr7 scaled \magstep1 \font\ninei=cmmi7 scaled \magstep1
\font\ninesy=cmsy7 scaled \magstep1 
\font\eightrm=cmr7 scaled 1140 
 
\font\sevenbf=cmbx7 \font\sevenrm=cmr7 \font\seveni=cmmi7
\font\sevensy=cmsy7 

\catcode`@=11
\newskip\ttglue
\newfam\ssfam

\def\fourteenpoint{\def\rm{\fam0\fourteenrm}
\textfont0=\fourteenrm \scriptfont0=\tenrm \scriptscriptfont0=\sevenrm
\textfont1=\fourteeni \scriptfont1=\teni \scriptscriptfont1=\seveni
\textfont2=\fourteensy \scriptfont2=\tensy \scriptscriptfont2=\sevensy
\textfont3=\fourteenex \scriptfont3=\fourteenex \scriptscriptfont3=\fourteenex
\def\it{\fam\itfam\fourteenit} \textfont\itfam=\fourteenit
\def\sl{\fam\slfam\fourteensl} \textfont\slfam=\fourteensl
\def\bf{\fam\bffam\fourteenbf} \textfont\bffam=\fourteenbf
\scriptfont\bffam=\tenbf \scriptscriptfont\bffam=\sevenbf
\def\tt{\fam\ttfam\fourteentt} \textfont\ttfam=\fourteentt
\def\ss{\fam\ssfam\fourteenss} \textfont\ssfam=\fourteenss
\tt \ttglue=.5em plus .25em minus .15em
\normalbaselineskip=16pt
\abovedisplayskip=16pt plus 4pt minus 12pt
\belowdisplayskip=16pt plus 4pt minus 12pt
\abovedisplayshortskip=0pt plus 4pt
\belowdisplayshortskip=9pt plus 4pt minus 6pt
\parskip=5pt plus 1.5pt
\setbox\strutbox=\hbox{\vrule height12pt depth5pt width0pt}
\let\sc=\tenrm
\let\big=\fourteenbig \normalbaselines\rm}
\def\fourteenbig#1{{\hbox{$\left#1\vbox to12pt{}\right.\n@space$}}}

\def\twelvepoint{\def\rm{\fam0\twelverm}
\textfont0=\twelverm \scriptfont0=\ninerm \scriptscriptfont0=\sevenrm
\textfont1=\twelvei \scriptfont1=\ninei \scriptscriptfont1=\seveni
\textfont2=\twelvesy \scriptfont2=\ninesy \scriptscriptfont2=\sevensy
\textfont3=\twelveex \scriptfont3=\twelveex \scriptscriptfont3=\twelveex
\def\it{\fam\itfam\twelveit} \textfont\itfam=\twelveit
\def\sl{\fam\slfam\twelvesl} \textfont\slfam=\twelvesl
\def\bf{\fam\bffam\twelvebf} \textfont\bffam=\twelvebf
\scriptfont\bffam=\ninebf \scriptscriptfont\bffam=\sevenbf
\def\tt{\fam\ttfam\twelvett} \textfont\ttfam=\twelvett
\def\ss{\fam\ssfam\twelvess} \textfont\ssfam=\twelvess
\tt \ttglue=.5em plus .25em minus .15em
\normalbaselineskip=14pt
\abovedisplayskip=14pt plus 3pt minus 10pt
\belowdisplayskip=14pt plus 3pt minus 10pt
\abovedisplayshortskip=0pt plus 3pt
\belowdisplayshortskip=8pt plus 3pt minus 5pt
\parskip=3pt plus 1.5pt
\setbox\strutbox=\hbox{\vrule height10pt depth4pt width0pt}
\let\sc=\ninerm
\let\big=\twelvebig \normalbaselines\rm}
\def\twelvebig#1{{\hbox{$\left#1\vbox to10pt{}\right.\n@space$}}}

\def\tenpoint{\def\rm{\fam0\tenrm}
\textfont0=\tenrm \scriptfont0=\sevenrm \scriptscriptfont0=\fiverm
\textfont1=\teni \scriptfont1=\seveni \scriptscriptfont1=\fivei
\textfont2=\tensy \scriptfont2=\sevensy \scriptscriptfont2=\fivesy
\textfont3=\tenex \scriptfont3=\tenex \scriptscriptfont3=\tenex
\def\it{\fam\itfam\tenit} \textfont\itfam=\tenit
\def\sl{\fam\slfam\tensl} \textfont\slfam=\tensl
\def\bf{\fam\bffam\tenbf} \textfont\bffam=\tenbf
\scriptfont\bffam=\sevenbf \scriptscriptfont\bffam=\fivebf
\def\tt{\fam\ttfam\tentt} \textfont\ttfam=\tentt
\def\ss{\fam\ssfam\tenss} \textfont\ssfam=\tenss
\tt \ttglue=.5em plus .25em minus .15em
\normalbaselineskip=12pt
\abovedisplayskip=12pt plus 3pt minus 9pt
\belowdisplayskip=12pt plus 3pt minus 9pt
\abovedisplayshortskip=0pt plus 3pt
\belowdisplayshortskip=7pt plus 3pt minus 4pt
\parskip=0.0pt plus 1.0pt
\setbox\strutbox=\hbox{\vrule height8.5pt depth3.5pt width0pt}
\let\sc=\eightrm
\let\big=\tenbig \normalbaselines\rm}
\def\tenbig#1{{\hbox{$\left#1\vbox to8.5pt{}\right.\n@space$}}}
\let\rawfootnote=\footnote \def\footnote#1#2{{\rm\parskip=0pt\rawfootnote{#1}
{#2\hfill\vrule height 0pt depth 6pt width 0pt}}}

\def\tenfoot{\tenpoint\hskip-\parindent\hskip-.1cm}

\overfullrule=0pt
\twelvepoint
\def\sbullet{\raise.2em\hbox{$\scriptscriptstyle\bullet$}}
\nofirstpagenotwelve
\hsize=16.5 truecm
\baselineskip 15pt

\def\ft#1#2{{\textstyle{{#1}\over{#2}}}}

\def\tc{\widetilde c}

\def\tV{\widetilde V}

\def\del{\partial}
\def\pb{\overline\psi}
\def\nott#1{{\mkern 4mu{{\mkern-4mu{\not}}} #1}}
\def\.{\,\,,\,\,}
\def\FF#1#2#3#4#5{\,\sb{#1}F\sb{\!#2}\!\left[\,{{#3}\atop{#4}}\,;{#5}\,\right]}

\def\cramp{\medmuskip = -2mu plus 1mu minus 2mu}
\def\uncramp{\medmuskip = 4mu plus 2mu minus 4mu}

\oneandahalfspace
\rightline{CTP TAMU--6/92}
\rightline{January 1992}

\vskip 2truecm
\centerline{\bf $SL(\infty,R)$ Kac-Moody symmetry of $W_\infty$ gravity}
\vskip 1.5truecm
\centerline{C.N. Pope\footnote{$^\star$}{\tenfoot Supported in part by the
U.S. Department of Energy, under
grant DE-FG05-91ER40633.}\footnote{}{\tenfoot Contribution to the
proceedings of the Trieste Summer Workshop on High-Energy Physics, Trieste,\nl
\indent$\,$ August 1991.}}

\vskip 1.5truecm

\centerline{\it Center for Theoretical Physics, Texas A\&M University,}
\centerline{\it College Station, TX 77843--4242, USA.}

\vskip 1.5truecm
\AB\singlespace
     Two-dimensional gravity in the light-cone gauge was shown by Polyakov
to exhibit an underlying $SL(2,R)$ Kac-Moody symmetry, which may be used to
express the energy-momentum tensor for the metric component $h_{++}$ in
terms of the $SL(2,R)$ currents {\it via}\ the Sugawara construction.  We
review some recent results which show that in a similar manner, $W_\infty$
and $W_{1+\infty}$ gravities have underlying $SL(\infty,R)$ and
$GL(\infty,R)$ Kac-Moody symmetries respectively.
\AE\oneandahalfspace

\np
\noindent
{\bf 1. Introduction}
\bigskip

     Einstein gravity, with action $\int {\sqrt g}R\, d^n x$, becomes trivial
in $n=2$ dimensions because the integrand is then a total derivative.  In
fact the action is then proportional to the Gauss-Bonnet expression for the
topological invariant $\chi$ -- the Euler number.  To circumvent this
problem, Polyakov proposed that one could construct a theory of
two-dimensional gravity with dynamics by taking the action to be given
instead as the induced quantum effective action for a non-critical matter
system in a curved two-dimensional background [1].  In the
light-cone gauge, where the dynamics of the metric is described just by its
$h_{++}$ component, this formulation of two-dimensional gravity has a
``hidden'' $SL(2,R)$ Kac-Moody symmetry, which enables one to write the
$h_{++}$ gauge field in terms of $SL(2,R)$ Kac-Moody currents [1].  The
energy-momentum tensor for $h_{++}$ may then be written in terms of the
Sugawara energy-momentum tensor for these currents [2].

     In this paper, we review some recent results in which the above
construction is generalised to induced $W_\infty$ and $W_{1+\infty}$
gravities.  One can show that in light-cone gauge, these theories exhibit
underlying $SL(\infty,R)$ and $GL(\infty,R)$ symmetries respectively [3].  For
technical reasons, which will be explained later, it is easier to describe
the $W_{1+\infty}$ case first.  We shall then indicate briefly how the
construction goes for $W_\infty$ gravity.

     The details of the derivation of the Kac-Moody symmetries for
$W_{1+\infty}$ and $W_\infty$ gravity are necessarily rather involved, because
of the complexity of the algebras.  Since the full details may be found in [3],
we shall concentrate here on the basic ideas, and the main results.  The method
used in [3] for deriving these results closely parallels the approach
used in [1] to derive the $SL(2,R)$ symmetry of two-dimensional gravity.  In
this paper, we therefore review the two-dimensional gravity calculation in some
detail, and then indicate the key steps in the generalisation to $W_{1+\infty}$
and $W_\infty$ gravity.  A complete description may be found in [3].

\bigskip\bigskip
\noindent{\bf 2. The $SL(2,R)$ symmetry of two-dimensional gravity}
\bigskip

     The induced Polyakov action for two-dimensional gravity is obtained as
the quantum effective action for a matter system in a general curved
two-dimensional background.  Since we shall be interested in gravity in the
light-cone gauge, we may impose this gauge choice at the outset and start by
considering a gauged classical matter Lagrangian of the form
$$
L=L_{\rm mat} +h\, T_{\rm mat}.\eqno(1)
$$
Here $L_{\rm mat}$ denotes the Lagrangian for the matter fields; $T_{\rm
mat}$ is their energy-momentum tensor; and $h$ denotes the $h_{++}$
component of the metric in the light-cone gauge.  The Lagrangian (1) can be
thought of as describing the gauging of the Virasoro algebra, realised on
the matter fields in $L_{\rm mat}$, that is generated by the current $T_{\rm
mat}$.  Under the residual diffeomorphisms preserving the light-cone gauge,
the gauge field $h$ transforms as
$$
\delta h=\bar\del k +k\, \del h-h\, \del k,\eqno(2)
$$
and the matter fields transform in the standard way dictated by their
tensorial structure.  The action given by (1) is then classically invariant
under these local gauge transformations.

     More or less any matter system that is classically conformally
invariant may be chosen in (1) in order to obtain
the induced two-dimensional gravity theory.  A suitable choice would be to
take the matter to be a complex fermion $\psi$, with $L_{\rm mat}=\pb
\bar\del \psi$, and $T_{\rm mat}=\ft12 \del \pb\, \psi-\ft12 \pb\,
\del\psi$. If the matter is now quantised then the quantum effective action
$\Gamma[h]$ may be defined as
$$
e^{-\Gamma[h]}\equiv \int D\psi\, e^{-{1\over \pi} \int L}.\eqno(3)
$$
This effective action transforms anomalously under the diffeomorphisms (2),
as may be seen by deriving the appropriate Ward identity.  We do this by
observing that
$$
{\delta \Gamma\over\delta h(z)}={1\over\pi}e^\Gamma\int D\psi\,
T(z)e^{-{1\over\pi}\int L} ,\eqno(4)
$$
and hence if we take the $\bar\del_z$ derivative the only contributions will
come from terms where it acts on the singularities in the operator-product
expansion of $T(z)$ with the $T(w)$ terms in the exponential.  Thus we have
$$
\bar\del {\delta \Gamma\over\delta h(z)}=-{1\over\pi^2} e^\Gamma\bar
\del\int D\psi\,
\Big({\del T(w)\over z-w}+{2T(w)\over(z-w)^2} +{c/2\over (z-w)^4}\Big)
e^{-{1\over\pi}\int L}.\eqno(5)
$$
Using $\bar\del(z-w)^{-1}=\pi\delta^{(2)}(z-w)$, it follows that
$$
\bar\del {\delta\Gamma\over\delta h}-h \del{\delta\Gamma\over\delta h}
-2\del h {\delta\Gamma\over\delta h}={c\over12\pi}\del^3 h.\eqno(6)
$$
Multiplying by a diffeomorphism parameter $k$ and integrating,
we see that this implies that under (2), $\Gamma[h]$ transforms
anomalously, as
$$
\delta\Gamma[h]=-{c\over12\pi}\int k\,\del^3 h.\eqno(7)
$$

     Viewing $\Gamma[h]$ as a (non-local) action for the light-cone metric
component $h$, we may now quantise $h$ itself, and calculate its correlation
functions.  Consider the $(N+1)$-point function
$$
\big\langle h(z) h(x_1)\cdots h(x_N)\big\rangle \equiv \int Dh \,
e^{-\Gamma[h]} \,h(z) h(x_1)\cdots h(x_N).\eqno(8)
$$
Acting on this with $-{c\over12}\del^3_z$, and using the anomalous Ward
identity (6), one obtains [1]
$$
\eqalign{
{c\over12\pi}\del^3_z&\big\langle h(z) h_1\cdots h_N\big\rangle =
\sum_{r=1}^N\Big\{\bar\del\delta^{(2)}(z-x_r)\big\langle h_1\cdots \nott{h_r}
\cdots h_N\big\rangle\cr
&\qquad-\del\delta^{(2)}(z-x_r)\big\langle h(z) h_1\cdots\nott{h_r} \cdots
h_N\big\rangle -2\delta^{(2)}(z-x_r)\big\langle \del h h1\cdots \nott{h_r}
\cdots  h_N\big\rangle\Big\},\cr}\eqno(9)
$$
where $h_r$ denotes $h(x_r)$, and we have used the result, proven by
functional integration by parts, that for any operator ${\cal O}$ depending
on $h$, $\big\langle \big(\delta\Gamma/\delta h\big){\cal O} \big\rangle
=\big\langle \delta{\cal O}/\delta h\big\rangle$.  The symbol $\nott{h_r}$
indicates that this particular gauge field is omitted in the product.  From the
identity  $\del_z^3 {(z-x_r)^2\over \bar z-\bar x_r}=2\pi \delta^{(2)}(z-x_r)$,
one may  integrate (9) to obtain, after a rescaling $h\rightarrow (6/c)h$ [1],
$$ \eqalign{
\big\langle h(z)h(x_1)\cdots h(x_r)\big\rangle&= \sum_{r=1}^N\Big\{
-{c\over6}{(z-x_r)^2\over(\bar z-\bar x_r)^2}\big\langle h(x_1)\cdots
\nott{h(x_r)}\cdots h(x_N)\big\rangle\cr
&-\Big[{2(z-x_r)\over \bar z-\bar x_r}+{(z-x_r)^2\over\bar z-\bar x_r}
\del_{x_r}\Big]\big\langle h(x_1)\cdots h(x_N)\big\rangle\Big\}.\cr}
\eqno(10)
$$

     Equation (10) is a recursion relation that gives the $(N+1)$-point
correlation function in terms of the $N$-point and $(N-1)$-point correlation
functions.  Starting from $\big\langle 1\big\rangle=1$ and $\big\langle
h(z)\big\rangle=0$, one may solve iteratively for the higher-point
functions.  For example, the two-point and three-point functions are given
by
$$
\eqalignno{
\big\langle h(x)h(y)\big\rangle&=-{c\over6}{(x-y)^2\over(\bar x-\bar y)^2},
&(11)\cr
\big\langle h(x)h(y)h(z)\big\rangle&={c\over3}{(x-y)(y-z)(z-x)\over (\bar
x-\bar y)(\bar y-\bar z)(\bar z-\bar x)}.&(12)\cr}
$$

    The condition on $h$ such that there is no anomaly in
the variation of $\Gamma[h]$ given by (7) is
$$
\del^3 h=0.\eqno(13)
$$
This equation may equivalently be obtained as the equation of motion
following from the induced action $\Gamma[h]$.  Its general
solution may be written as
$$
h(z,\bar z)=J^{(1)}(\bar z) -2 J^{(0)}(\bar z) z + J^{(-1)}(\bar z) z^2.
\eqno(14)
$$
Using this expansion, one may re-express (10) as a recursion relation for
the correlation functions of the coefficient operators $J^i$.  The result is
[1]
$$
\eqalign{
\big\langle J^{i}(\bar z)J^{j_1}(\bar x_1)\cdots J^{j_N}(\bar x_N)
\big\rangle&=\sum_{r=1}^N\Big\{-{c\over12} {\delta^i_{j_r}\over (\bar z-\bar
x_r)^2} \big\langle J^{j_1}(\bar x_1)\cdots\nott{J^{j_r}(\bar x_r)}\cdots
J^{j_N}(\bar x_N)\big\rangle\cr
& -{f^{ij_r}{}_{k_r}\over\bar z-\bar x_r} \big\langle
J^{k_r}(\bar x_r) J^{j_1}(\bar x_1)\cdots \nott{J^{j_r}(\bar x_r)} \cdots
J^{j_N}(\bar x_N)\big\rangle\},\cr}\eqno(15)
$$
where $\eta^{ij}$ is the inverse Cartan-Killing metric and $f^{ij}{}_k$ are
the structure constants of $SL(2,R)$.  The two-point and three-point
correlation functions (11) and (12) become
$$
\eqalignno{
\big\langle J^i(\bar x)J^j(\bar y)\big\rangle&=-{\eta^{ij}\over(\bar x-\bar
y)^2},&(16)\cr
\big\langle J^i(\bar x)J^j(\bar y)J^k(\bar z)\big\rangle&= {f^{ijk}\over (\bar
x-\bar y)(\bar y-\bar z)(\bar z-\bar x)}.&(17)\cr}
$$
These correlation functions, and the recursion relation, are precisely those
for the currents of an $SL(2,R)$ Kac-Moody algebra, arising from an
$SL(2,R)$ WZW model.

\bigskip\bigskip
\noindent{\bf 3. The $GL(\infty,R)$ symmetry of $W_{1+\infty}$ gravity}
\bigskip

     The generalisation from two-dimensional light-cone gravity to
$W_{1+\infty}$ gravity is made by introducing additional gauge fields $A_i$ for
all the higher-spin currents of $W_{1+\infty}$, and replacing the Lagrangian
(1)
by $$
L=L_{\rm mat} +\sum_{i=-1}^\infty A_i \tV^i.\eqno(18)
$$
Here $\tV^i(z)$ is the spin-$(i+2)$ current of $W_{1+\infty}$, and $A_i$ is its
corresponding gauge field.  Details of the $W_\infty$ and $W_{1+\infty}$
algebras may be found in [4], and also in the lectures on $W$ algebras and $W$
gravity in this volume [5].  We are adopting the notation of [5] here, and
using
$\tV^i$ to denote the currents of $W_{1+\infty}$, and $V^i$ to denote those for
$W_\infty$.

     The operator-product expansions for the $W_{1+\infty}$ currents take the
form
$$
\tV^i(z)\tV^j(w)\ \sim\ -\sum_{\ell\ge0} {\widetilde
f}^{ij}_{2\ell}\bigl(\partial_z,
\partial_w\bigr){\tV^{i+j-2\ell}(w)\over{z-w}}\
-\tc_i\delta^{ij}(\partial_z)^{2i+3}{1\over{z-w}},\eqno(19)
$$
with
$$
\widetilde f_{2\ell}^{ij}(m,n)={1\over 2(2\ell+1)!}
{\widetilde\phi}^{ij}_{2\ell}\ M^{ij}_{2\ell}(m,n),\eqno(20)
$$
where
$$
{\widetilde\phi}^{ij}_{2\ell}=\FF43{\ft12\. \ft12\. -\ell-\ft12\. -\ell}{
-i-\ft12\. -j-\ft12\. i+j-2\ell +\ft52}1\eqno(21)
$$
and
$$
M^{ij}_{2\ell}(m,n)=\sum_{k=0}^{2\ell+1}(-)^k{2\ell+1\choose k} (2i-2\ell+2)_k
[2j+2-k]_{2\ell+1-k}\, m^{2\ell+1-k} n^k.\eqno(22)
$$
The central charges $\tc_i$ are given by
$$
{\widetilde c}_i={2^{2i-2}((i+1)!)^2\over (2i+1)!!(2i+3)!!}c.\eqno(23)
$$
(Fuller details may be found in [4,5].)

     It is clear that the same steps as those described in section 2 for pure
gravity may be followed here for $W_{1+\infty}$ gravity.  The analogue of the
anomalous Ward identity (6) turns out to be
$$
\bar\partial {\delta\Gamma\over \delta A_i} +\sum_{\ell\ge0}
\widetilde f^{ij}_{2\ell}(\partial,-\partial_A)\Big( {\delta \Gamma\over \delta
A_{i+j-2\ell}} A_j\Big)={\tc_i\over\pi} \partial^{2i+3} A_i,\eqno(24)
$$
where the derivative $\del$ in $\widetilde f^{ij}_{2\ell}(\del,-\del_A)$ acts
on
everything to its right, whilst $\del_A$ acts only on the explicit $A_j$ term
in the following parentheses.  The analogue of the diffeomorphism
transformation (2) is
$$
\delta A_i=\bar\partial k_i +\sum_{\ell\ge0}
\sum_{j=-1}^{i+2\ell+1}\widetilde
f^{j,i-j+2\ell}_{2\ell}(\partial_A,\partial_k)A_j\, k_{i-j+2\ell},\eqno(25)
$$
where $\del_A$ acts only on $A_j$, and $\del_k$ acts only on the transformation
parameters $k_{i-j+2\ell}$.  From (24) and (25), it follows that under a
spin-$i+2$ transformation with parameter $k_i$, the effective action
$\Gamma[A]$ has an anomalous variation
$$
\delta_{k_i}\Gamma[A]=-{\tc_i\over\pi}\del^{2i+3}A_i.\eqno(26)
$$

     The anomalous Ward identity (24) may be used to obtain a recursion
relation for the correlation functions for the gauge fields of quantum
$W_{1+\infty}$ gravity.  Repeating the steps described in section 2 for pure
gravity, we find [3]
$$
\eqalign{
\big\langle A_i(z)A_{j_1}(x_1)\cdots& A_{j_N}(x_N)\big\rangle =\cr
 &-\sum_{r=1}^N \tc_{j_r}
(2j_r+2)! \delta^{i, j_r}{(z-x_r)^{2i+2}\over (\bar z-{\bar x}_r)^2}
\big\langle
A_{j_1}(x_1)\cdots \nott{A}_{j_r}(x_r)\cdots A_{j_N}(x_N)\big\rangle\cr
&-\sum_{k\ge -1} \sum_{\ell=0}^{[(i+k)/2]} \sum_{r=1}^N {\tc_{j_r}(2j_r+2)!
\over \tc_k
(2k+2)!}\delta_{i+k-2\ell,j_r} \widetilde f^{ik}_{2\ell}(\partial_z,
-\partial_{A_k})\cr
&\times {(z-x_r)^{2i+2}\over (\bar z-{\bar x}_r)} \big\langle A_k(x_r)
A_{j_1}(x_1)\cdots \nott{A}_{j_r}(x_r)\cdots A_{j_N}(x_N)\big\rangle,\cr}
\eqno(27)
$$
after a rescaling of the gauge fields $A_i$ according to
$$
A_i\rightarrow {1\over \tc_i(2i+2)!}A_i.\eqno(28)
$$
This recursion relation may be solved iteratively for the correlation
functions for $W_{1+\infty}$ gravity.  For example, we find [3]
$$
\big\langle A_i(x,\bar x)A_j(y,\bar y)\big\rangle= \tc_i(2i+2)!\,
\delta_{ij}\, {(x-y)^{2i+2}\over (\bar x-\bar y)^2}\eqno(29)
$$
for the two-point function.  For the three-point function, we find [3]
$$
\big\langle A_i(x,\bar x) A_j(y,\bar y) A_k(z, \bar z)\big\rangle =
\widetilde N_{ijk} {(x-y)^{i+j-k+1}(y-z)^{j+k-i+1} (z-x)^{k+i-j+1}\over (\bar
x-\bar y)(\bar y-\bar z)(\bar z-\bar x)},\eqno(30)
$$
when $i+j+k$ is even, and zero when $i+j+k$ is odd.  $\widetilde N_{ijk}$ is
defined by
$$
\widetilde N_{ijk}\equiv {(2i+2)!(2j+2)!(2k+2)!\over
(i+j-k+1)!(j+k-i+1)!(k+i-j+1)!}\widetilde P_{ijk},\eqno(31)
$$
with $\widetilde P_{ijk}$  given by
$$
\widetilde P_{ijk}=\ft12 \tc_k \widetilde\phi^{ij}_{i+j-k}.\eqno(32)
$$
$\widetilde P_{ijk}$ is manifestly symmetric in $i$ and $j$.  Although it is
not
manifest, it is in fact totally symmetric in $i$, $j$ and $k$ [3].

     From (26), we see that the conditions on the gauge fields for the
transformation rules to be anomaly free are
$$
\del^{2i+3}A_i=0,\eqno(33)
$$
generalising (13) for pure gravity.  Thus we may expand the spin-$(i+2)$ gauge
field $A_i$ as a polynomial of degree $(2i+3)$ in $z$:
$$
A_i(z,\bar z)=\sum_{m=-i-1}^{i+1}  {2i+2\choose i+1+m}
J^i_m(\bar z) (-z)^{i+1+m}.\eqno(34)
$$
Substituting this into the two-point function (29) and three-point function
(30)  for the gauge fields $A$, we obtain the two-point and three-point
functions for the ``expansion coefficients'' $J^i_m(\bar z)$. For the two-point
function, we find [3]
$$
\big\langle J^i_m(\bar x)J^j_n(\bar y)\big\rangle= {\widetilde K^{ij}_{mn}\over
(\bar x-\bar y)^2},\eqno(35)
$$
where the $\widetilde K^{ij}_{mn}$ are given by
$$
\widetilde K^{ij}_{mn}= (-1)^{i+1+m}\,
\tc_i(i+1+m)!(i+1-m)!\, \delta^{ij}\delta_{m+n,0}.\eqno(36)
$$
After some algebra, we find that the three-point function for $J^i_m$ can be
written as [3]
$$
\big\langle J^i_m(\bar x)J^j_n(\bar y)J^k_p(\bar z)\big\rangle
={\widetilde Q^{ijk}_{mnp} \over (\bar x-\bar y)(\bar y-\bar z)(\bar z-\bar
x)},\eqno(37)
$$
when $i+j+k$ is even, and zero when $i+j+k$ is odd.  The coefficients
$\widetilde Q^{ijk}_{mnp}$ are given by
\cramp
$$
\eqalign{
&\widetilde Q^{ijk}_{mnp} = \delta_{m+n+p,0}\times\cr
&\times\sum_{q\ge0}{(i+1+m)!(i+1-m)!(j+1+n)!(j+1-n)!
(k+1+p)!(k+1-p)!\widetilde P_{ijk}(-)^{j+1-m+p+q}\over (j-k-m+q)!(i+1+m-q)!
(j-i+p+q)! (k+1-p-q)! (k+i-j+1-q)!q!}\cr}\eqno(38)
$$
\uncramp

     By analogy with the pure-gravity case described in section 2, we should
expect that $\widetilde K^{ij}_{mn}$ in the two-point function (35) and
$\widetilde Q^{ijk}_{mnp}$ in the three-point function (37) should be related
to
the Cartan-Killing metric and structure constants of the Lie algebra from some
underlying Kac-Moody symmetry of $W_{1+\infty}$ gravity.  This indeed turns out
to be the case; the Lie algebra in question here is $GL(\infty,R)$ [3].  It can
in fact be described as the algebra generated by the ``wedge'' of
$W_{1+\infty}$
generators $\tV^i_m$ [4](where $m$ is the Laurent-mode index in the expansion
of
the spin-$(i+2)$ $W_{1+\infty}$ current $V^i(z)$), with $m$ is restricted to
lie
in the range
$$
-(i+1)\le m\le (i+1),\qquad\qquad i\ge -1.\eqno(39)
$$
Thus we may think of the ``expansion coefficients'' $J^i_m(\bar z)$ in (34) as
being $GL(\infty,R)$ Kac-Moody currents, with the underlying Lie algebra
being described by the subset of $W_{1+\infty}$ generators $X^i_m=\tV^i_m$
specified by (39).

     Because $GL(\infty,R)$ is infinite dimensional some care has to be taken
when defining the Cartan-Killing metric, since a naive definition such as ${\rm
tr}\big(X^i_m X^j_n\big)$ would diverge when the trace over
infinite-dimensional matrices is taken.  The way around this is to observe that
the trace operation is simply a procedure for projecting onto the singlet
term in the product of the two adjoint representations generated by $X^i_m$ and
$X^j_n$.  Thus if we can find a way of recognising the singlet term
{\it without} taking the trace then we can read off the Cartan-Killing metric
as the coefficient of the singlet, thereby avoiding the divergent coefficient
that it would acquire if we were to take the trace.  There is indeed a simple
way to do this for the $GL(\infty,R)$ algebra described by the subset of
$W_{1+\infty}$ generators $\tV^i_m$ specified by (39) [3].  The method exploits
the fact that one can define a product operation for the generators of
$W_{1+\infty}$.  (This ``lone-star'' product $\tV^i_m\star \tV^j_n$ is
associative, which implies that it defines a Lie bracket $[\tV^i_m,\tV^j_n]
=\tV^i_m\star \tV^j_n - \tV^j_n\star \tV^i_m$ that automatically satisfies the
Jacobi identity.)  The zero-mode of the spin-1 current, $V^{-1}_0$, turns out
to commute with every generator in $W_{1+\infty}$.  Thus it behaves like the
identity operator, and is a singlet under $GL(\infty,R)$.  The Cartan-Killing
metric for $GL(\infty,R)$ may therefore be defined as the coefficient of
$X^{-1}_0$ in the lone-star product $X^i_m\star X^j_n$ of $GL(\infty,R)$
generators.

     As shown in [3], the Cartan-Killing metric $\widetilde\eta^{ij}_{mn}$ for
$GL(\infty,R)$, defined by the procedure described above, is related to the
coefficients $\widetilde K^{ij}_{mn}$ appearing in (35) by $\widetilde
K^{ij}_{mn}=\ft18\widetilde\eta^{ij}_{mn}$.  After some algebra, one can show
by substituting the expansion (34) into the recursion relation (27) that the
coefficients $J^i_m(\bar z)$ satisfy the recursion relation [3]
$$
\eqalign{
&\big\langle J^i_n(\bar z) J^{j_1}_{m_1}({\bar x}_1) \cdots J^{j_N}_{m_N}({\bar
x}_N) \big\rangle=
\ft18 \sum_r {\eta^{i j_r}_{n m_r}\over (\bar z-{\bar x}_r)^2} \big\langle
J^{j_1}_{m_1}({ \bar x}_1)\cdots \not J^{j_r}_{m_r}({\bar x}_r) \cdots
J^{j_N}_{m_N}({\bar x}_N) \big\rangle \cr
&\qquad\qquad\qquad-\sum_r \sum_{k\ge0} \ \sum_{m_k=-k-1}^{k+1}
{f^{i j_r m_k}_{n m_r k}\over(\bar
z-{\bar x}_r)}
\big\langle J^k_{m_k}({\bar x}_r) J^{j_1}_{m_1}({\bar x}_1)\cdots \nott{
J}^{j_r}_{m_r}({\bar x}_r)\cdots J^{j_N}_{m_N}({\bar x}_N)
\big\rangle,\cr}\eqno(40)
$$
where $f^{ijk}_{mnp}$ are the structure constants of $GL(\infty,R)$, read off
from the wedge subalgebra of $W_{1+\infty}$.  Equation (40) is precisely the
recursion relation for the Kac-Moody currents from a $GL(\infty,R)$ WZW model.
Thus, we have established that $W_{1+\infty}$ gravity has an underlying
$GL(\infty,R)$ Kac-Moody symmetry.

\bigskip\bigskip
\noindent{\bf 4. Conclusions}
\bigskip

     We have seen in the previous section that Polyakov's result for the
underlying $SL(2,R)$ Kac-Moody symmetry of two-dimensiuonal gravity generalises
to a $GL(\infty,R)$ symmetry for $W_{1+\infty}$ gravity.  The case of
$W_\infty$ gravity can be handled in a very similar way.  There is one further
complication here, resulting from the fact that the corresponding ``lone-star''
product for $W_\infty$ does not generate a spin-1 term on the right-hand side,
and so one cannot identify the Cartan-Killing metric for the wedge subalgebra
(which is $SL(\infty,R)$ in this case) as the coefficient of $V^{-1}_0$.  The
solution to this problem is to view $W_\infty$ as a special case of a
one-parameter family of parametrisations of the $W_{1+\infty}$ algebra [3,4].
The $W_\infty$ algebra, for which the spin-1 generator can be truncated from
the
$W_{1+\infty}$ algebra, can then be approached {\it via} a limiting procedure,
in which one rescales generators so as to retain a non-zero coefficient for
$V^{-1}_0$ in the $W_\infty$ limit.  The details are described in [3].  The
conclusion is that $W_\infty$ gravity has an underlying $SL(\infty,R)$
Kac-Moody symmetry.

     In the work of [2], the $SL(2,R)$ Kac-Moody symmetry was exploited for
studying aspects of two-dimensional quantum gravity.  In particular, the
energy-momentum tensor for the gauge field $h$, which must be added to the
Lagrangian to preserve general covariance, was expressed in terms of the
$SL(2,R)$ Kac-Moody currents by using the Sugawara construction.  In principle,
a similar procedure should be possible for $W_\infty$ or $W_{1+\infty}$
gravity.  Indeed a generalisation of the Sugawara construction, known as
Casimir algebras,  has been given for $W$-extended conformal algebras [6].
Essentially, one builds the higher-spin currents by using symmetric invariant
tensors of the Lie algebra underlying the $W$ algebra.  For example, $W_N$ can
be viewed as the Casimir algebra of $SU(N)$, and the higher-spin currents can
be built in terms of the $SU(N)$ Kac-Moody currents $J^A(z)$ by using the
invariant $d$ tensors of $SU(N)$.  Thus we have $T=\eta_{AB}:J^AJ^B:$,
$W=d_{ABC}:J^A J^B J^C:$, {\it etc}.  The procedure described in section 3 for
extracting the Cartan-Killing metric for $GL(\infty,R)$ can be
straightforwardly
extended to obtain its arbitrary-rank symmetric $d$ tensors.  The limiting
procedure described above can be used to obtain the analogous tensors for
$SL(\infty,R)$.  Thus in principle it should be possible to repeat the steps of
[2], and express the energy-momentum tensor for the gauge fields of $W_\infty$
or $W_{1+\infty}$ gravity in terms of the corresponding $SL(\infty,R)$ or
$GL(\infty,R)$ Kac-Moody currents.  The main outstanding problem seems to be
that one obtains divergent results that would presumably have to be regularised
in some way [3]; there are indications from other considerations that such a
regularisation ought to be possible [7].  It may be that achieving a better
understanding of how to do this will require finding a higher-dimensional
interpretation for the infinite set of higher-spin currents of $W_\infty$.

\bigskip
\centerline{\bf Acknowledgments}
\bigskip

     I am very grateful to my collaborators in the work described in this
review, namely Shawn Shen, Kaiwen Xu and Kajia Yuan, and to the organisers
of the Trieste Summer School in High-Energy Physics for hospitality.
\bigskip\bigskip

\singlespace
\centerline{\bf REFERENCES}
\frenchspacing
\bigskip

\item{[1]} A.M. Polyakov, Mod. Phys. Lett. {\bf A2} (1987) 893.

\item{[2]} V. Knizhnik,  A.M. Polyakov and A.B. Zamolodchikov, Mod. Phys.
Lett. {\bf A3} (1988) 819.

\item{[3]} C.N. Pope, X. Shen, K.W. Xu and K. Yuan, ``$SL(\infty,R)$ symmetry
of quantum $W_\infty$ gravity,'' preprint, CTP TAMU-37/91,
Imperial/TP/90-91/29; to appear in Nucl. Phys. B.

\item{[4]} C.N. Pope, L.J. Romans and X. Shen, Phys. Lett. {\bf 236B}
(1990) 173;\nl
Nucl. Phys. {\bf B339} (1990) 191;\nl
Phys. Lett. {\bf 242B} (1990) 401.

\item{[5]} C.N. Pope, ``Lectures on $W$ algebras and $W$ gravity,'' in the
Proceedings of the 1991 Summer School on High-Energy Physics, Trieste, 1991
(this volume).

 \item{[6]} F. Bais, P. Bouwknegt, M. Surridge and K. Schoutens, Nucl. Phys.
{\bf B304} (1988) 348; 371.

\item{[7]} C.N. Pope, L.J. Romans and X. Shen, Phys. Lett. {\bf 254B} (1991)
401.

\end